\documentclass[aps,prb,twocolumn,showpacs]{revtex4-1}
\usepackage{mathrsfs, amsfonts, amsmath, amssymb, graphicx, bm}
\usepackage[usenames,dvipsnames,svgnames,table]{xcolor}
\usepackage[colorlinks=true,linkcolor=RoyalBlue,citecolor=RoyalBlue]{hyperref}

\begin{document}
\title{Spin-Mechanical Inertia in Antiferromagnet}

\author{Ran Cheng}
\affiliation{Department of Physics, Carnegie Mellon University, Pittsburgh, PA 15213}

\author{Xiaochuan Wu}
\affiliation{Department of Physics, Carnegie Mellon University, Pittsburgh, PA 15213}

\author{Di Xiao} 
\affiliation{Department of Physics, Carnegie Mellon University, Pittsburgh, PA 15213}

\begin{abstract}
Angular momentum conservation has served as a guiding principle in the interplay between spin dynamics and mechanical rotations. However, in an antiferromagnet with vanishing magnetization, new fundamental rules are required to properly describe spin-mechanical phenomena. Here we show that the N\'{e}el order dynamics affects the mechanical motion of a rigid body by modifying its inertia tensor in the presence of strong magnetocrystalline anisotropy. This effect depends on temperature when magnon excitations are considered. Such a spin-mechanical inertia can produce measurable consequences at small scales.
\end{abstract}


\maketitle

\section{Introduction}

Spin-mechanics, also known as magnetomechanics, is a venerable branch of modern physics that has attracted continuous attention. It explores the coupled dynamics of quantum spins and mechanical motions of a crystal background,~\cite{ref:Chud_conserve,ref:review} where the conservation of angular momentum serves as the governing principle. For example, when a paramagnet is placed in a magnetic field to polarize the atomic spins inside, it undergoes a spontaneous rotation to balance the angular momentum acquired from the magnetic field, known as the Einstein-de Haas effect.~\cite{ref:EdH} The reverse process, \textit{i.e.}, a crystal rotation generating spin polarization, has also been discovered around the same period by Barnett.~\cite{ref:Barnett}

Different from paramagnets, spins in a ferromagnetic material order collectively into a magnetization even in the absence of an external magnetic field. The magnetization serves as an order parameter and carries an intrinsic angular momentum. As a consequence of the angular momentum conservation, a reorientation of the order parameter is necessarily accompanied by a mechanical rotation, and \textit{vice versa}. Following the seminal discovery of the Einstein-de Haas effect, the idea that angular momentum can transfer between magnetic and mechanical degrees of freedom has fertilized a broad spectrum of applications such as magnetic force microscopy~\cite{ref:AFM,ref:cantilever1,ref:cantilever2}, mechanical manipulations of spins~\cite{ref:phononlaser,ref:nanomechanics,ref:Maekawa}, \textit{etc}.

However, this simple picture seems to break down in antiferromagnets (AFs) with vanishing magnetization. In this case, the ground state is characterized by the N\'{e}el order parameter which does not carry an angular momentum. Only when the N\'{e}el order is driven into motion does a small magnetization develop;~\cite{ref:AFMR,ref:spinpump} it is this induced magnetization that is subjected to angular momentum conservation. In other words, unlike its ferromagnetic counterparts, the order parameter dynamics in AFs is not directly dictated by any conservation law. Therefore, to study how spin-mechanical effects can manifest through the N\'{e}el order parameter instead of the small magnetization, one must seek new physics beyond angular momentum conservation.

In this regard, a well-established phenomenon provides a critical hint: The coordinated motion of antiparallel magnetic moments in an AF creates a fictitious inertia in the effective N\'{e}el order dynamics~\cite{ref:Haldane,ref:DW,ref:Rasing,ref:Gomonay}, in sharp contrast to the non-inertial behavior of the magnetization dynamics in ferromagnets: The N\'{e}el order behaves more like a massive particle that can be accelerated via external forces rather than an angular momentum regulated directly by the conservation law. Regarding this unique feature, it is tempting to ask whether the fictitious inertia of the N\'{e}el order can lead to any mechanical consequence.

In this paper, we demonstrate that the fictitious inertia of the N\'{e}el order modifies the inertia tensor that characterizes the rigid background rotation, giving rise to a mechanically measurable effect. We term this effect spin-mechanical inertia, which reflects a quantum correction to the otherwise classically defined inertia. Our claim is justified by modeling a collinear AF as a hybrid system consisting of antiferromagnetic spins and a rigid mechanical rotation; they couple through an easy-axis anisotropy. When the two subsystems operate at vastly different time scales, their coupled motion can be solved by the adiabatic approximation. The spin-mechanical inertia is then derived as a result of the adiabatic approximation. Furthermore, by considering magnon excitations, we find that the spin-mechanical inertia is subject to a reduction in two aspects: zero-point quantum fluctuation and thermal fluctuations. The former is independent of temperature $T$, whereas the latter results in an appreciable temperature dependence when the thermal energy $k_BT$ is comparable to the magnon gap. Finally, we invoke the path integral formalism to derive a criterion for the adiabatic approximation by inquiring into non-adiabatic corrections, which turns out to be well suppressed in typical situations. Our result establishes spin-mechanical inertia as an essential ingredient of spin-mechanics in the context of AFs.

This paper is organized as follows. In Sec.~\ref{macro}, we study the simplest case of a uniform AF to illustrate the essential physics, supplemented by a discussion on possible detection schemes. In Sec.~\ref{magnon}, we consider the temperature dependence of spin-mechanical inertia arising from magnon excitations. In Sec.~\ref{sec:non-adiabatic}, non-adiabatic corrections are derived using the path integral formalism. In Sec.~\ref{discussion}, we discuss the underlying physics of spin-mechanical inertia from several fundamental aspects and potential issues that may complicate out result. Mathematical details are presented in the Appendices.

\section{Macrospin Model}\label{macro}

To capture the essential physics, we first consider the simplest case where the antiferromagnetic ordering is spatially homogeneous and described by two macrospins $\bm{S}_A$ and $\bm{S}_B$ with equal magnitude $|\bm{S}_A|=|\bm{S}_B|=S$. The two macrospins couple through the Heisenberg exchange interaction $H=J\bm{S}_A\cdot\bm{S}_B$. Defining the N\'{e}el order as $\bm{N}=(\bm{S}_A-\bm{S}_B)/2S$, we follow the standard procedure to eliminate the small magnetization $\bm{m}=\bm{S}_A+\bm{S}_B$~\cite{ref:Soliton,ref:Fradkin,ref:Auerbach}, which yields an effective action of $\bm{N}$:
\begin{align}
  \mathcal{S}_N&=\frac{\hbar^2V}{2ZJa^3}\int \mathrm{d}t \left|\partial_{t}\bm{N}\right|^2, \label{eq:N}
\end{align}
where $V$ is the system volume, $a$ is the lattice constant, and $Z$ is the coordination number. For simplicity, we have assumed a cubic lattice. In fact, $\mathcal{S}_N$ is equivalent to the action of a rigid rod with supporting point on its center of mass. This property indicates that the N\'{e}el order acquires an effective inertia from the exchange interaction between $\bm{S}_A$ and $\bm{S}_B$.

\begin{figure}[t]
	\includegraphics[width=\columnwidth]{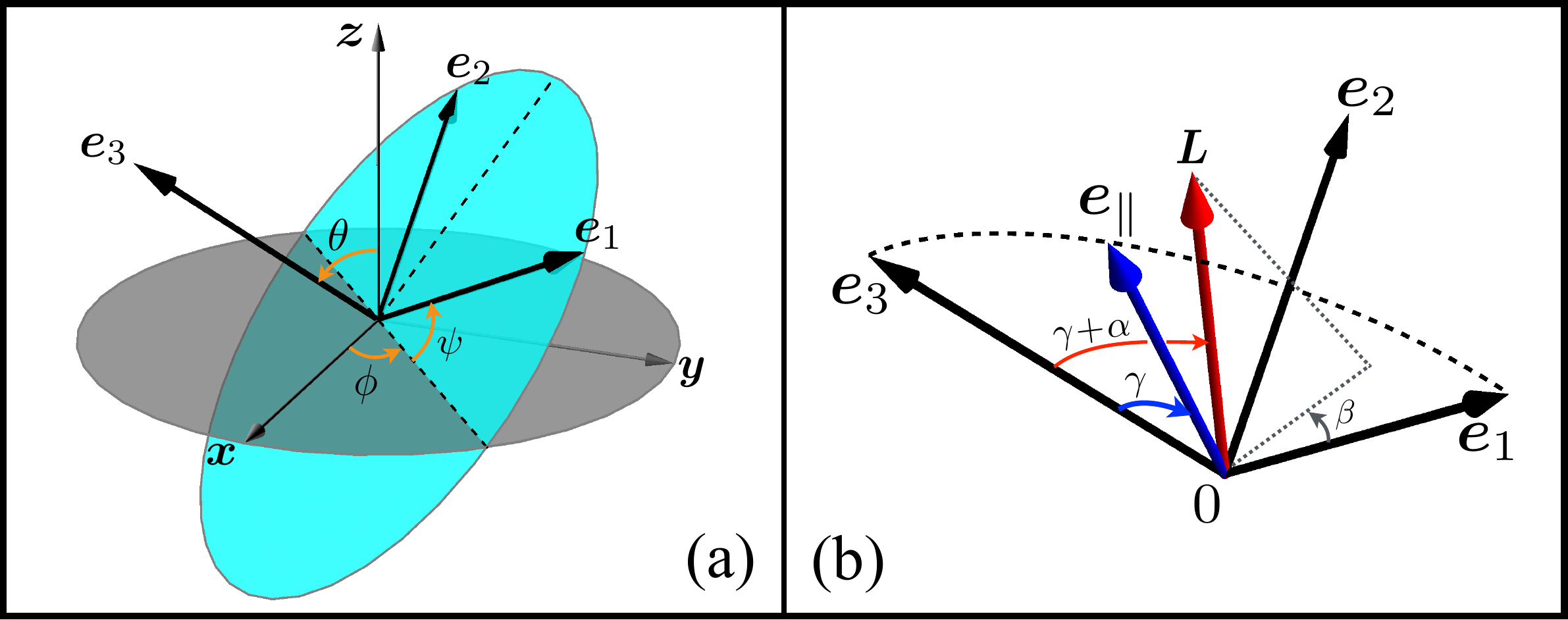}
	\caption{(a) Euler angles $\theta$, $\phi$, and $\psi$ specify the relative orientation of the body frame labeled by the principal axes $\bm{e}_1$-$\bm{e}_2$-$\bm{e}_3$ with respect to the laboratory frame $x$-$y$-$z$. (b) The $\bm{e}_{\parallel}$ and $\bm{L}$ vectors in the body frame. }
	\label{fig:angles}
\end{figure}

Next we picture the crystal background as a rigid body. When it rotates about a fixed-axis, its kinetic energy is $T=\frac12I\dot{\varphi}^2$ with $I$ the moment of inertia and $\varphi$ the angle of rotation around the axis. More generally, when the rigid body rotates about a fixed point (a spinning top), its instantaneous orientation is characterized by the principal axes of the body-frame, which are specified by three Euler angles $\bm{\lambda}(t)\equiv\{ \theta(t),\phi(t),\psi(t) \}$ as shown in Fig.~\ref{fig:angles}(a). We now consider a symmetric top in which the $\bm{e}_1$ and $\bm{e}_2$ axes are equivalent and the system preserves cylindrical symmetry with respect to the $\bm{e}_3$ axis. Accordingly, the kinetic energy is $T=\frac12I_{\perp}(\dot{\theta}^2+\sin^2\theta\dot{\phi}^2)+\frac12I_{\parallel}(\dot{\psi}+\cos\theta\dot{\phi})^2$, where $I_{\perp}$ and $I_{\parallel}$ are the moments of inertia with respect to the $\mathbf{e}_1$ (or $\mathbf{e}_2$) and $\mathbf{e}_3$ axes, respectively~\cite{ref:Greiner}. In the absence of gravitational torques (known as the Euler top), the system action only has the kinetic energy, thus its action can be written as~\cite{ref:pathint}
\begin{align}
 \mathcal{S}_R&=\frac12\int \mathrm{d}t\, G_{ij}(\lambda)\dot{\lambda}^i\dot{\lambda}^j, \label{eq:RB}
\end{align}
where repeated indices are summed. $G_{ij}(\lambda)$ made up by $I_{\perp}$, $I_{\parallel}$, and $\bm{\lambda}$; it plays the role of an effective metric in the parameter space spanned by the Euler angles.

We assume that the spin subsystem couple to the rigid crystal background (an Euler top) through an easy-axis anisotropy, described by the action
\begin{align}
 \mathcal{S}_K=\frac{KV}{2a^3}\int \mathrm{d}t \left| \bm{N}\cdot\bm{e}_{\parallel}(\bm{\lambda}) \right|^2, \label{eq:anisotropy}
\end{align}
where $K>0$ in our convention~\cite{note_anisotropy}. The easy-axis $\bm{e}_{\parallel}$ is a function of $\bm\lambda$ since its direction depends on the orientation of the rigid body. It is this $\lambda$-dependence that connects the two subsystems. If the anisotropy $K$ is sufficiently strong such that the N\'{e}el order is able to adjust to the easy-axis at any instant of time, then the entire system moves as a whole as if a rigid rod is firmly attached to the Euler top, which defines an adiabatic motion of the hybrid system.

We now check the behavior of the hybrid system in the adiabatic limit $\hbar\omega_{rb}/K\rightarrow0$, where $\omega_{rb}$ is the rotational frequency of the rigid body. Deviations from the adiabatic limit (\textit{i.e.} non-adiabatic corrections) will be discussed in Sec.~\ref{sec:non-adiabatic}. In this limit, $\mathcal{S}_K$ is a constant, so the effective action becomes $\mathcal{S}_{\rm eff}=\mathcal{S}_N+\mathcal{S}_R$. As illustrated in Fig.~\ref{fig:angles}(b), we fix the body frame by choosing the $\bm{e}_1$ axis coplanar with $\bm{e}_3$ and $\bm{e}_{\parallel}$: $\bm{e}_{\parallel}=\cos\gamma\bm{e}_3+\sin\gamma\bm{e}_1$. Such a choice is always possible for a symmetric Euler top. The effective action $\mathcal{S}_{\rm eff}$ then becomes
\begin{align}
	\mathcal{S}_{\rm eff}[\bm{\lambda}]=\frac12\int \mathrm{d}t\left[ G_{ij}+\frac{\hbar^2V}{ZJa^3}g_{ij} \right]\dot{\lambda}^i\dot{\lambda}^j, \label{eq:Seff}
\end{align}
where $g_{ij}$ is a correction of the parameter-space metric tensor originating from the N\'{e}el order dynamics; it is a function of the Euler angles and $\gamma$:
\begin{align}
	g_{\theta\theta}=&\left(3+\cos2\gamma-2\sin^{2}\gamma\cos2\psi\right)/4, \notag\\
	g_{\phi\phi}=&\sin^{2}\gamma\cos^{2}\psi+\left(\cos\gamma\sin\theta-\sin\gamma\cos\theta\sin\psi\right)^2, \notag\\
	g_{\psi\psi}=&\sin^{2}\gamma, \notag\\
	g_{\theta\phi}=&\sin\gamma\cos\psi\left(\cos\gamma\cos\theta+\sin\gamma\sin\theta\sin\psi\right), \notag\\
	g_{\phi\psi}=&\sin\gamma\left(\sin\gamma\cos\theta-\cos\gamma\sin\theta\sin\psi\right), \notag\\
	g_{\theta\psi}=&\sin\gamma\cos\gamma\cos\psi. \label{eq:g}
\end{align}
In the presence of $g_{ij}$, the inertia tensor is no longer diagonal in the body-frame labeled by the principal axes. Or equivalently, we can say that the principal axes themselves are changed by the spin-mechanical coupling. We term this effect \textit{spin-mechanical inertia}. In the limit that $\gamma\rightarrow0$, \textit{i.e.}, when the easy-axis $\bm{e}_{\parallel}$ is parallel to the principal axis $\bm{e}_3$, only two components survive: $g_{\theta\theta}=1$ and $g_{\phi\phi}=\sin^2\theta$. In this case, $g_{ij}$ reduces to a spherical metric and the principal axes do not change. Nevertheless, moments of inertia associated with the principal axes, $I_{\perp}$ and $I_{\parallel}$, are still modified.

In Eq.~\eqref{eq:Seff}, the strength of spin-mechanical inertia seems to be proportional to the system volume $V$. However, since $G_{ij}$ scales as $Vd^2$ with $d$ the body dimension transverse to the instantaneous rotation axis, the relative strength of spin-mechanical inertia scales as $d^{-2}$ instead of $V$. Therefore, we expect a pronounced effect only in small systems.

\subsection{Rotation about a fixed-axis}

\begin{figure}[t]
	\includegraphics[width=\columnwidth]{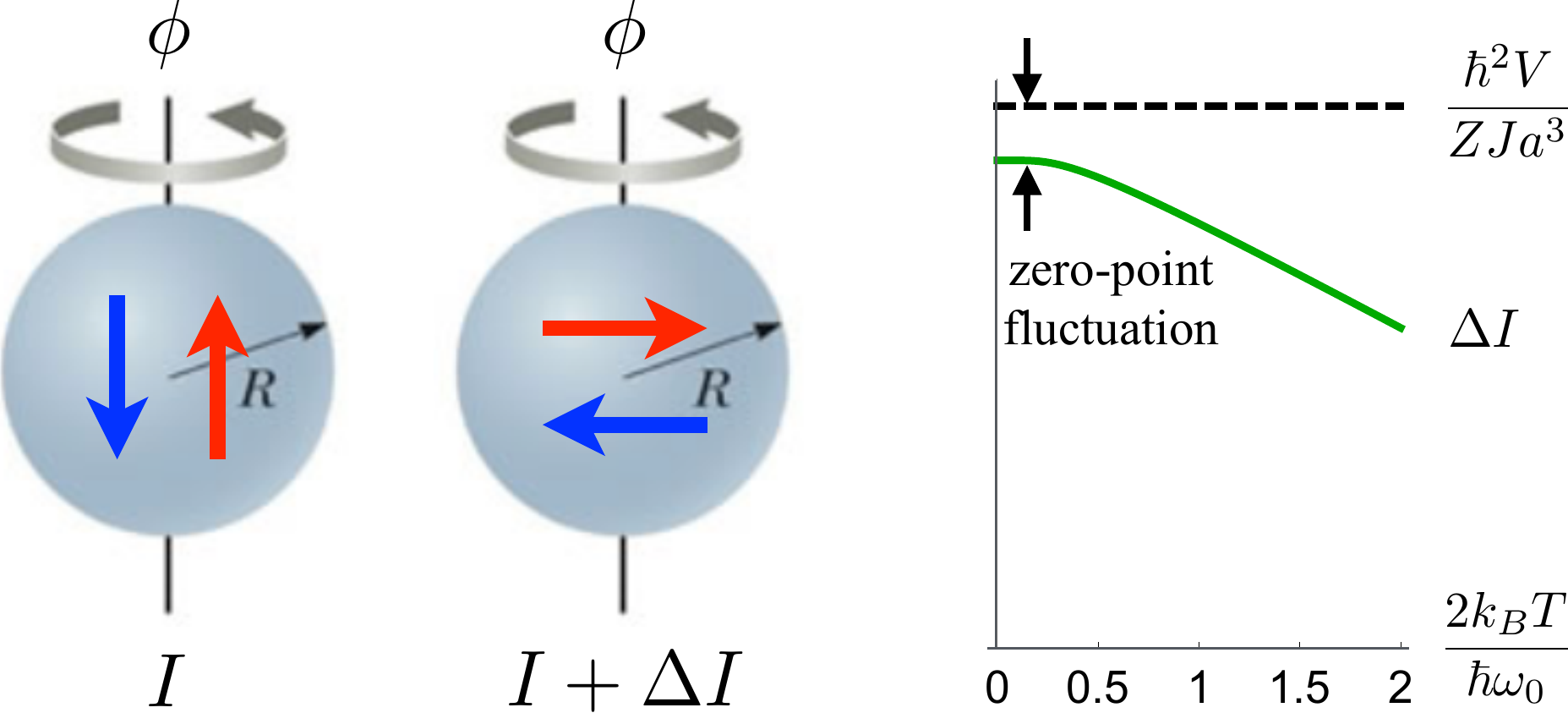}
	\caption{Left: Schematics of the moment of spin-mechanical inertia $\Delta I$ in a sphere with uniform mass distribution. Right: When averaged over magnon excitations, $\Delta I$ decreases with an increasing temperature, supplemented by a residual zero-temperature reduction due to quantum fluctuation. Parameters for the plot: $V=10^3a^3$ and $J=10^2\hbar\omega_0$.}
	\label{fig:rotation}
\end{figure}

To demonstrate the physical consequences of the spin-mechanical inertia, we now consider a rotation about a fixed axis, where the inertia tensor reduces to a moment of inertia $I$. For instance, a sphere with uniform mass distribution has a constant $I$ regardless of how the sphere is suspended. By contrast, the moment of spin-mechanical inertia depends on the relative orientation of $\bm{N}$ with respect to the rotation axis $\hat{\bm{z}}$. According to Eq.~\eqref{eq:g}, $\Delta I=\hbar^2V/(ZJa^3)|\hat{\bm{z}}\times\bm{N}|^2$. $\Delta I$ reaches maximum for $N\perp\hat{\bm{z}}$, and thus the period of oscillation measured by a torsion balance, as illustrated in Fig.~\ref{fig:rotation}, reaches a maximum for $N\perp\hat{\bm{z}}$. The relative correction $\Delta I/I$, which scales as $d^{-2}$ as mentioned above, gets larger when the sphere gets smaller.

Admittedly, the torsion balance shown in Fig.~\ref{fig:rotation} is probably not a realistic setup for detection at microscopic scales.  For example, if the sphere considered above refers to an AF molecule, then both the spin dynamics and the molecular rotation should be treated quantum mechanically. Therefore, a possible way to observe the spin-mechanical inertia is to measure the change of the rotational spectrum when a N\'{e}el ordering is introduced (\textit{e.g.}, by lowering the temperature). For example, if we regard the AF molecule as a quantum rotor with moment of inertia $I$, the energy is quantized as $E=\hbar^2n^2/2I$ with $n=0,1,2\cdots$. Since the spin-mechanical inertia changes $I$ into $I+\Delta I$, the energy splitting is slightly reduced. By monitoring the shift of spectral lines stemming from transitions between the ground state and states with large $n$ (so that the change is magnified by $n^2$), one should be able to identify the existence of spin-mechanical inertia.

We estimate the effect in an AF molecule consisting of thousands of atoms. Suppose the magnetic moments originate from transition metal elements, such as iron and nickel, and take $J$ to be tens of meV (similar to the superexchange interaction in bulk antiferromagnetic crystals), then $\Delta I/I$ falls somewhere between $10^{-2}$ and $10^{-3}$. If we further consider that the value of $J$ in magnetic molecules is smaller than that in magnetic crystals, then the spin-mechanical inertia should be more significant than the above estimation.

Here we emphasize one point: The validity of the rigid body action Eq.~\eqref{eq:RB} does not require that the mechanical motion is classical. In fact, Eq.~\eqref{eq:RB} can describe a fully quantum rotation in the path integral formalism to be exploited below.  The spin-mechanical correction of the inertia tensor holds, whether the mechanical motion is classical or quantum.

\subsection{Rotation about a fixed-point}

\begin{figure}[t]
	\includegraphics[width=\columnwidth]{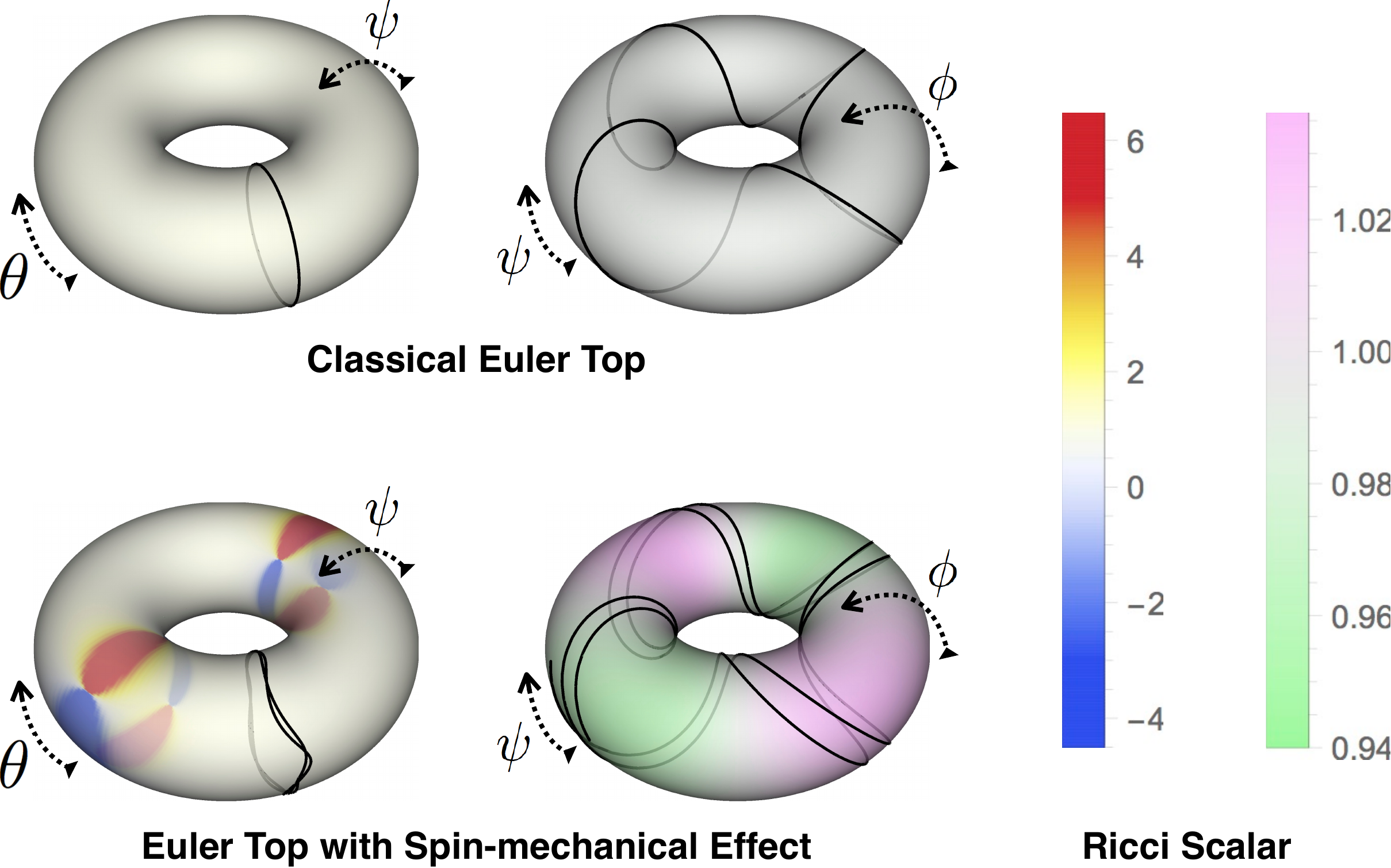}
	\caption{Ricci curvature (colored) and geodesic curves (solid black) in the parameter space of the Euler angles. The Euler top has $I_{\perp}=I_{\parallel}/2=10\hbar^2V/(ZJa^3)$ and $\bm{e}_{\parallel}\perp\bm{e}_3$. Plot range: $\theta\in[0,\pi)$, $\phi\in[0,2\pi)$, and $\psi\in[0,2\pi)$. In the absence (presence) of spin-mechanical inertia, the Ricci curvature is a constant $R_0=1$ (a periodic function in all Euler angles). $R$ diverges at $\theta=0,\pi$ and $\psi=\pm\pi/2,\pm\pi$, where the Euler angles are ill-defined. In the $\theta-\psi$ subspace, we cut off the color bar at $R=R_0\pm5.5$.}
	\label{fig:curvature}
\end{figure}

For rotations about a fixed point, the tensorial nature of inertia is reinstated. Although this general case is not necessary for practical measurements, it entails a beautiful geometrical interpretation of spin-mechanical inertia. To this end, it is adequate to consider a classical Euler top, the motion of which can be obtained by minimizing the effective action Eq.~\eqref{eq:Seff} with respect to the Euler angles $\bm{\lambda}$. The result is a geodesic equation in the parameter space:
\begin{align}
 \ddot{\lambda}^k+\Gamma^{k}_{ij}\dot{\lambda}^i\dot{\lambda}^j=0, \label{eq:geodesic}
\end{align}
where $\Gamma^{k}_{ij}=\frac12\mathcal{G}^{k\ell}(\partial_i\mathcal{G}_{j\ell}+\partial_j\mathcal{G}_{i\ell}-\partial_{\ell}\mathcal{G}_{ij})$ is the connection with $\mathcal{G}_{ij}=G_{ij}+\frac{\hbar^2V}{ZJa^3}g_{ij}$ the total metric and $\mathcal{G}^{ij}$ satisfying $\mathcal{G}^{i\ell}\mathcal{G}_{\ell j}=\delta^i_j$. The metric tensor $\mathcal{G}_{ij}$ fully determines the geometry of the parameter space.

In the absence of spin-mechanical effect, the geodesic curve solved by Eq.~\eqref{eq:geodesic} corresponds to a trivial great arc in the $\theta-\psi$ subspace as shown in Fig.~\ref{fig:curvature}. Here $\theta$ is a constant of motion while $\phi$ and $\psi$ precess uniformly. We have chosen parameters to make the $\psi-\phi$ phase portrait commensurate. To further understand this feature from a geometrical perspective, we calculate the Ricci curvature of the parameter space defined as
\begin{align}
 R=\mathcal{G}^{ij}\left(\frac{\partial\Gamma^k_{ij}}{\partial\lambda^k}-\frac{\partial\Gamma^k_{ik}}{\partial\lambda^j} +\Gamma^{\ell}_{ij}\Gamma^k_{k\ell}-\Gamma^{\ell}_{ik}\Gamma^k_{j\ell}\right),
\end{align}
which is analogous to the inverse radius of a sphere. It turns out that $R$ is a constant throughout the parameter space if spin-mechanical inertia is disregarded, which explains why the geodesic curve is trivial. However, the spin-mechanical inertia introduces an induced metric $g_{ij}$ on top of $G_{ij}$, which changes the geometry and distorts the parameter space so that the Ricci curvature is no longer a constant. Consequently, the geodesic curve characterizing the rigid body rotation deflects from its original path, as if a fictitious gravity appears in the parameter space. As depicted in Fig.~\ref{fig:curvature}, a free Euler top has no nutation and the geodesic curve projected onto the $\psi-\phi$ subspace is commensurate with our chosen parameter. The spin-mechanical inertia breaks these properties: It not only generates a small nutation, but also decommensurates the orbit in the $\psi-\phi$ subspace.

\section{Magnon Excitations}\label{magnon}

So far the N\'{e}el order $\bm{N}$ has been regarded as a uniform vector, which is a reasonable approximation at low temperatures. To investigate how the spin-mechanical inertia is affected by thermal fluctuations embedded in the spin dynamics, we need to go beyond the macrospin model and introduce magnon excitations. Since magnons are inhomogeneous deviations from the uniform ground state, we need to generalize the N\'{e}el vector into a staggered field $\bm{N}=\bm{N}(t,\bm{r})$, and promote Eq.~\eqref{eq:N} into the nonlinear sigma model~\cite{ref:Haldane}
\begin{align}
  \mathcal{S}_N=\frac{\hbar^2}{2ZJa^3}\int \mathrm{d}^4r\left(\eta^{\mu\nu}\partial_\mu\bm{N}\cdot\partial_\nu\bm{N}\right), \label{eq:NLSM}
\end{align}
where $c=Ja/\hbar$ is the spin wave velocity, $r^\mu=\{t,\bm{r}\}$ is the joint spacetime coordinate, $\mathrm{d}^4r=\mathrm{d}t\mathrm{d}^3\bm{r}$, and $\eta^{\mu\nu}=\rm{diag}[1,-c^2,-c^2,-c^2]$ is the spacetime metric of the laboratory frame. The anisotropy term is now
\begin{align}
 \mathcal{S}_K=\frac{K}{2a^3}\int \mathrm{d}^4r|\bm{N}(t,\bm{r})\cdot\bm{e}_{\parallel}(\bm\lambda)|^2. \label{eq:SK}
\end{align}
Next we employ the standard procedure~\cite{ref:Auerbach} to decompose the staggered field into
\begin{align}
	\bm{N}(t,\bm{r})=\bm{L}(t)\sqrt{1-|\boldsymbol{\pi}(t,\bm{r})|^2}+\bm{\pi}(t,\bm{r}), \label{eq:decomp}
\end{align}
where $\bm{L}(t)$ is a time-dependent unit vector and $\bm{\pi}(t,\bm{r})$ is the magnon field (a transverse fluctuation) that satisfies $\bm{L}\cdot\bm{\pi}=0$. We restrict our discussion to the low-temperature regime where $|\bm{\pi}|\ll1$. Our goal is to eliminate $\bm{\pi}$ and derive an effective action of $\bm{L}$ with a temperature-dependent coefficient, which is supposed to replace the original action Eq.~\eqref{eq:N}.

To this end, we insert Eq.~\eqref{eq:decomp} into Eq.~\eqref{eq:NLSM} and Eq.~\eqref{eq:SK}. After some tedious algebra, as detailed in Appendix~\ref{ave}, we obtain
\begin{align}
	\mathcal{S}_N=&\frac{\hbar^2}{2ZJa^3} \int \mathrm{d}^4r\Big\{ (1-|\bm{\pi}|^2)\partial_t\bm{L}\cdot\partial_t\bm{L} \notag\\
	&\quad +\eta^{\mu\nu}\left[ \partial_\mu\bm{\pi}\cdot\partial_\nu\bm{\pi}+\frac{(\bm{\pi}\cdot\partial_\mu\bm{\pi})(\bm{\pi}\cdot\partial_\nu\bm{\pi})}{1-|\bm{\pi}|^2} \right] \notag\\
	&\quad +\eta^{\mu\nu}\pi^a\pi^b\partial_\mu \mathfrak{e}_a\cdot\partial_\nu\mathfrak{e}_b \Big\}, \label{eq:SN}\\
	\mathcal{S}_K=&\frac{K}{2a^3}\int \mathrm{d}^4r\left[ (1-|\bm{\pi}|^2)\left|\bm{L}\cdot\bm{e}_{\parallel}\right|^2+\left|\bm{\pi}\cdot\bm{e}_{\parallel}\right|^2 \right], \label{eq:SA}
\end{align}
where $\{\mathfrak{e}_a\}$ forms a set of local coordinates labeling the transverse plane normal to the instantaneous $\bm{L}(t)$; the magnon field $\bm{\pi}(t,\bm{r})$ resides in this plane. In Eqs.~\eqref{eq:SN} and \eqref{eq:SA}, terms that will not survive the thermal averaging operation below have been omitted; they are listed in Appendix~\ref{ave}. The sum $\mathcal{S}_N+\mathcal{S}_K$ defines a coupled field theory consisting of $\bm{L}$ and $\bm{\pi}$.

Integrating out the $\bm{\pi}$ field, again, requires the adiabatic approximation. But at finite temperatures, the meaning of the adiabatic approximation changes. It now means that the rigid body rotates sufficiently slow such that the staggered field remains in thermal equilibrium with respect to the instantaneous crystal orientation at all times. This allows us to take a thermal average over the $\bm{\pi}$ field by freezing $\bm{L}(t)$, which finally leads to an effective description of $\bm{L}(t)$ with temperature-dependent parameters. The result comes in the form of a thermally-averaged action $\bar{\mathcal{S}}_{L}\equiv\langle \mathcal{S}_N+\mathcal{S}_K \rangle^{\rm th}$. As derived in Appendix~\ref{ave}, $\bar{\mathcal{S}}_{L}$ reads
\begin{align}
	\bar{\mathcal{S}}_{L}=\frac{V}2\int \mathrm{d}t\left[\frac{\hbar^2}{ZJa^3}\Theta(T) \dot{\bm{L}}^{2}+\frac{K}{a^3}\left(\bm{L}\cdot\bm{e}_{\parallel}\right)^{2}\right], \label{eq:Leff}
\end{align}
where the temperature-dependent factor is
\begin{align}
     \Theta(T)=1-\frac{Za^3}{V}\sum_{\bm k}\frac{J}{\hbar\omega_k}\coth\frac{\hbar\omega_k}{2k_BT}, \label{eq:IT}
\end{align}
with $k_B$ the Boltzmann constant and $\omega_k$ the dispersion. It is clear that the net effect of magnon excitations is to replace the original action Eq.~\eqref{eq:N} with Eq.~\eqref{eq:Leff}, in which the unit vector $\bm{L}(t)$ becomes an effective order parameter, and the coefficient acquires a temperature dependence through $\Theta(T)$. This change, in turn, yields a temperature-dependent spin-mechanical inertia reflected in a thermally averaged action
\begin{align}
 \bar{\mathcal{S}}_{\rm eff}[\bm{\lambda}]=\frac12\int \mathrm{d}t\left[ G_{ij}+\Theta(T)\frac{\hbar^2V}{ZJa^3}g_{ij} \right]\dot{\lambda}^i\dot{\lambda}^j, \label{eq:SeffT}
\end{align}
which replaces our previous result Eq.~\eqref{eq:Seff}.

To assess the significance of magnon excitations in the spin-mechanical inertia, we consider an AF nanomagnet with \textit{quantized} magnon modes subjected to rotations about a fixed-axis (as illustrated in Fig.~\ref{fig:rotation}). The moment of spin-mechanical inertia is simply
\begin{align}
 \Delta I(T)=\frac{\hbar^2V}{ZJa^3}\Theta(T) \label{eq:SMI}
\end{align}
with $\Theta(T)$ given by Eq.~\eqref{eq:IT}. Because of the prominent energy splitting at the nanometer scale, the lowest mode $\omega_0$ is well separated from all other modes. Since $\coth x$ converges to unity rapidly with $x$, the dominant contribution to the temperature dependence originates from the lowest mode, which scales as $(\frac{J}{\hbar\omega_{0}})(\frac{a^3}{V})\coth\frac{\hbar\omega_{0}}{2k_BT}$ according to Eq.~\eqref{eq:IT}.

Without loss of essential physics, we will only keep the lowest mode, which occurs at $\bm{k}=0$ with a gap $\hbar\omega_0$ being few Kelvins. In typical antiferromagnets such as MnF$_2$, the ratio $K/J$ is around $10^{-3}$ to $10^{-4}$. Since the N\'{e}el temperature is in a loose sense proportional to $J$ that far exceeds the gap $\hbar\omega_0\sim\sqrt{ZJK}\sim10^{-2}J$, it is still within the low temperature regime even when $k_BT$ is comparable to $\hbar\omega_0$. In MnF$_2$, for example, $\hbar\omega_0\sim2$ K while the N\'{e}el temperature is around 60 to 80 K, thus our theory remains valid up to few Kelvins. 

With these considerations, we plot the spin-mechanical inertia Eq.~\eqref{eq:SMI} as a function of temperature in Fig.~\ref{fig:angles}, assuming $V/a^3=10^3$ and $J/\hbar\omega_{0}=10^2$. There are two noticeable features in Fig.~\ref{fig:angles}: (i) $\Delta I(T)$ starts to bend down at around $k_BT\sim\hbar\omega_0$, which marks the onset of substantial thermal fluctuations. (ii) There is a residual reduction of spin-mechanical inertia even at zero temperature, $\Delta I(T\rightarrow0)<\frac{\hbar^2 V}{ZJa^3}$. This is attributed to the zero-point quantum fluctuation of $\bm{N}$ around the easy-axis. According to Eq.~\eqref{eq:IT}, $\hbar\omega_0\sim\sqrt{ZJK}$, this zero-temperature correction vanishes in the limit $K\rightarrow\infty$.

\section{Non-Adiabatic Effect}\label{sec:non-adiabatic}

We finally derive a criterion for the adiabatic assumption employed in previous sections. Since the influence of magnon excitations has been resolved by the temperature dependence of the spin-mechanical inertia during the thermal averaging operation, we can now treat $\bm{L}(t)$ as the real order parameter~\cite{note_temp}. In the body frame depicted by Fig.~\ref{fig:angles}(b), $\bm L$ can be decomposed as
\begin{align}
\bm{L}=\cos(\gamma+\alpha)\bm{e}_3+&\sin(\gamma+\alpha)\cos\beta\bm{e}_1 \notag\\
+&\sin(\gamma+\alpha)\sin\beta\bm{e}_2, \label{eq:alphabeta}
\end{align}
where $\alpha$ and $\beta$ are two independent variables characterizing the deviation of $\bm{L}$ from the easy-axis $\bm{e}_{\parallel}$. For large but finite anisotropy $K$, misalignment between $\bm{L}$ and $\bm{e}_{\parallel}$ should be small, so we assume that $\alpha\ll1$ and $\beta\ll1$. As detailed in Appendix~\ref{nA}, by adopting the path integral formalism~\cite{ref:Fradkin} and expanding the action $\bar{\mathcal{S}}_L$ up to second order in $\alpha$ and $\beta$, we can analytically integrate out the fast variable $\bm{L}$ as
\begin{align}
 \mathcal{Z}&=\int\mathcal{D}\bm{\lambda}\mathcal{D}\bm{L}\delta^3(\bm{L}^2-1)\exp\left[ \frac{i}{\hbar}\left(\mathcal{S}_{R}+\bar{\mathcal{S}}_L\right) \right] \notag\\
 &=\int\mathcal{D}\bm\lambda\exp\left[ \frac{i}{\hbar}\left( \bar{\mathcal{S}}_{\rm eff}+\Delta S \right) \right], \label{eq:Z}
\end{align}
where $\bar{\mathcal{S}}_{\rm eff}$ is given by Eq.~\eqref{eq:SeffT}. The $\Delta S$ term includes all non-adiabatic corrections, which can be expressed as a series summation
\begin{align}
 \Delta S=&\frac{\hbar^2V}{2ZJa^3}\Theta(T)\sum_{n=1}^{\infty} \Theta(T)^{n}(-1)^n \notag\\
 &\quad\times\int \mathrm{d}t X^{\rm T}(\bm{\lambda})\left[\frac{\hbar^2}{ZJK}
 \partial_t^2\right]^n X(\bm{\lambda}), \label{eq:expasion}
\end{align}
where the vector $X^{\rm T}(\bm{\lambda})=\{X_1(\bm{\lambda}),X_2(\bm{\lambda})\}$ represents a particular combination of the Euler angles:
$X_1(\bm{\lambda})=-\sin\psi\dot{\theta}+\cos\psi\sin\theta\dot{\phi}$ and $X_2(\bm{\lambda})=\sin\gamma(\cos\theta\dot{\phi}+\dot{\psi})-\cos\gamma(\cos\psi\dot{\theta}+\sin\theta\sin\psi\dot{\phi})$. In Eq.~\eqref{eq:expasion}, the small quantity of expansion is $\frac{\hbar^2}{ZJK}\partial_t^2$, which is proportional to $(\omega_{rb}/\omega_0)^2$ with $\omega_{rb}$ the frequency of rigid body rotation and $\hbar\omega_0$ the anisotropy gap used earlier. In typical AFs such as transition metal oxides or fluorides, $\omega_0$ is in the Terahertz regime, which coincides with the frequency scale of vibrational modes in a magnetic molecule. On the other hand, $\omega_{rb}$ corresponds to the frequency of rotational modes that is typically far below the vibrational frequency. Therefore, the adiabatic condition is likely to be well respected.

\section{Discussions}\label{discussion}

It is worthwhile to distinguish the spin-mechanical effect explored in this paper from the well-established magnetoelastic phenomena. The latter scenario primarily focuses on the hybridization of magnetic and mechanical excitations. For example, when spin dynamics is driven by a current, mechanical vibrations are agitated~\cite{ref:Gomonay2}. By contrast, our attention is paid on the ground state where the effect is maximum at zero temperature; elementary excitations reduce the strength of the effect. The spin-mechanical inertia we predict is a conceptual progress that poses a serious challenge to the common belief that moment of inertia is a classical quantity.

We also mention that the spin-mechanical inertia does not modify the inertial mass of the crystal. It only makes sense when a rigid body undergoes rotations instead of linear motions. By definition, the moment of inertia is the response coefficient of the angular acceleration versus an external torque. In classical mechanics, this coefficient turns out to be, but is not defined as, a quantity that is merely determined by the mass distribution of the body. What we have shown in this paper is that this response coefficient also depends on the spin degree of freedom of the constituent atoms in AFs, which cannot be described by classical mechanics.

Besides magnons, thermal excitations also come in the form of phonons. Phonons play a significant role in spin-mechanical effects of ferromagnets because local distortions of the lattice background directly couple to the magnetization in the form of $\dot{\bm m}\cdot(\bm \nabla \times \bm u)$, where $\bm{u}(t,\bm{r})$ is the local displacement field of the lattice and $\bm{m}(t,\bm{r})$ is the local magnetization vector. This form of coupling can either be justified by angular momentum conservation or derived from a simple model including the easy-axis anisotropy~\cite{ref:Chud_phonon}. In an AF, the latter approach is apparently more reasonable, since angular momentum conservation does not explicitly rule the N\'{e}el order dynamics. (Caution: Angular momentum is always conserved, but the N\'{e}el order does not carry one.)

It is straightforward to check that the local lattice distortion $\bm \nabla \times \bm u$ always couple to $\bm{m}(t,\bm{r})$ instead of $\bm{N}(t,\bm{r})$. However, in a collinear AF, local magnetization develops only when the staggered field is driven into motion~\cite{ref:AFMR,ref:spinpump}: $\bm{m}\sim\bm{N}\times\dot{\bm{N}}/J$. Therefore, the magnitude of $\bm{m}$ scales as $\sqrt{K/J}$, which is typically few percents. This implies that phonons are far less important in collinear AFs than in ferromagnets regarding spin-mechanical effects. Nevertheless, our discussions refer to transverse phonons only. There might be a strong effect from the longitudinal phonons as they would affect the exchange interaction $J$ by modulating distances between neighboring spins.

\begin{acknowledgments}
	We are grateful to A.~H.~MacDonald, J.~Zhu, and S.~Okamoto for inspiring discussions. X.W. also thanks N.~P.~Ong for insightful comments. This work was supported by the Department of Energy, Basic Energy Sciences, Grant No.~DE-SC0012509.  D.X. also acknowledges support from a Research Corporation for Science Advancement Cottrell Scholar Award.
	
	R.C. and X.W. contributed equally to this work.
\end{acknowledgments}

\appendix

\section{}\label{ave}

The time dependence of $\bm{L}\left(t\right)$ originates from the rigid body rotation, while that of $\bm{\pi}(t,\bm{r})$ stems from thermal agitations. Equation~\eqref{eq:decomp} can be further written as
\begin{align}
\bm{N}\left(t,\bm{r}\right) =\boldsymbol{L}\left(t\right)\sqrt{1-|\boldsymbol{\pi}|^{2}}+\pi^{a}\left(t,\bm{r}\right)\mathfrak{e}_{a}\left(t\right),
\end{align}
where $\left\{ \mathfrak{e}_{a}\left(t\right)\right\} $ for $a=1,2$ and $\mathfrak{e}_{0}\equiv\bm{L}$ together forms a local orthonormal base with respect to the instantaneous $\bm{L}(t)$, satisfying $\bm{L}\bm{L}+\sum_{a=1}^{2}\mathfrak{e}_{a}\mathfrak{e}_{a}=\mathbb{I}$. To improve visual clarity, hereafter we will omit the spacetime argument unless necessary. We define $\mathscr{A}_{a0}\equiv\mathfrak{e}_{a}\cdot\partial_{t}\bm{L}$, $\mathscr{A}_{0a}\equiv\bm{L}\cdot\partial_{t}\mathfrak{e}_{a}$, and $\mathscr{A}_{ab}\equiv\mathfrak{e}_{a}\cdot\partial_{t}\mathfrak{e}_{b}$ as temporal connections of the local base. They obey $\partial_{t}\mathfrak{e}_{A}=\mathfrak{e}_{B}\mathscr{A}_{BA}$, $\mathscr{A}_{AB}+\mathscr{A}_{BA}=0$, and $\mathscr{A}_{CA}\mathscr{A}_{CB}=\partial_{t}\mathfrak{e}_{A}\cdot\partial_{t}\mathfrak{e}_{B}$, where $A,B$ take $0,1,2$ (\textit{c.f.}, $a,b$ only take $1,2$) and repeated indices are summed. With these notations, we have 
\begin{align}
\partial_{\mu}\bm{N} & =\delta_{\mu0}\left(\partial_{t}\bm{L}\right)\sqrt{1-|\bm{\pi}|^{2}}+\frac{\bm{\pi}\cdot\partial_{\mu}\bm{\pi}}{\sqrt{1-|\boldsymbol{\pi}|^{2}}}\bm{L }\notag\\
&\qquad+(\partial_{\mu}\pi^{a})\mathfrak{e}_{a}+\delta_{\mu0}\mathfrak{e}_{a}\mathscr{A}_{ab}\pi^{b},\label{eq:partialN}
\end{align}
where $\mu=\left\{ 0,1,2,3\right\} \equiv\left\{ t,x,y,z\right\} $. Now we insert Eq.~\eqref{eq:partialN} into the actions Eq.~\eqref{eq:SN} and~\eqref{eq:SA}, and reorganize the terms of the total action into three parts
\begin{align}
\mathcal{S}_{AF}\equiv\mathcal{S}_{N}+\mathcal{S}_{A}=\int{\rm d}^{4}r\left(\mathcal{L}_{L}+\mathcal{L}_{\pi}+\mathcal{L}_{{\rm odd}}\right),\label{eq:threeterms}
\end{align}
where the first two terms are, respectively,
\begin{align}
\mathcal{L}_{L} & =\frac{\hbar^2}{2ZJa^3} \left[(1-|\boldsymbol{\pi}|^{2})\partial_{t}\bm{L}\cdot\partial_{t}\bm{L}+\partial_{t}\mathfrak{e}_{a}\cdot\partial_{t}\mathfrak{e}_{b}\pi^{a}\pi^{b}\right] \notag\\ &\qquad+\frac{K}{2a^{3}}\left(\bm{L}\cdot\bm{e}_{\parallel}\right)^{2},\label{eq:LN}\\
\mathcal{L}_{\pi} & =\frac{\hbar^2}{2ZJa^3} \eta^{\mu\nu}\mathscr{G}_{ab}\partial_{\mu}\pi^{a}\partial_{\nu}\pi^{b} \notag\\ &\qquad+\frac{K}{2a^{3}}\left[\left(\bm{\pi}\cdot\bm{e}_{\parallel}\right)^{2}-|\bm{\pi}|^{2}\left(\bm{L}\cdot\bm{e}_{\parallel}\right)^{2}\right],\label{eq:Lpi}
\end{align}
where $\mathscr{G}_{ab}$ is the metric in the local base~\cite{ref:Auerbach} defined as
\begin{align}
\mathscr{G}_{ab}\left(\bm{\pi}\right)=\delta_{ac}\frac{\pi^{c}\pi^{d}}{1-|\boldsymbol{\pi}|^{2}}\delta_{db}+\delta_{ab},
\end{align}
here $a,b,c,d$ take $1,2$. The third term of Eq.~\eqref{eq:threeterms} reads
\begin{align}
\mathcal{L}_{{\rm odd}}=&\frac{\hbar^2}{2ZJa^3} \Big[\mathscr{A}_{ab}\pi^{b}\partial_{t}\pi^{a} +\frac{\mathscr{A}_{a0}\pi^{a}\bm{\pi}\cdot\partial_{t}\bm{\pi}}{\sqrt{1-|\bm{\pi}|^{2}}} \notag\\
& +\sqrt{1-|\bm{\pi}|^{2}}\mathscr{A}_{a0}\left(\partial_{t}\pi^{a}+\mathscr{A}_{ab}\pi^{b}\right) \Big] \notag\\
& +\frac{K}{a^{3}}\left(\bm{L}\cdot\bm{e}_{\parallel}\right)\left(\mathfrak{e}_{a}\cdot\bm{e}_{\parallel}\right)\pi^{a}\sqrt{1-|\bm{\pi}|^{2}},\label{eq:odd}
\end{align}
which, to be shown below, will vanish identically under thermal averaging.

Next we integrate out the $\bm{\pi}$ field and derive an effective Lagrangian for $\bm{L}$. However, since $\bm{\pi}$ represents magnon excitations driven by thermal fluctuations, the integration should be performed in the Euclidean space where temperature plays the role of time. In other words, we are dealing with an adiabatic process in which $\boldsymbol{\pi}$ stays in thermal equilibrium with respect to the instantaneous $\bm{L}$. Retaining to the non-interacting order in the small $\bm{\pi}$ field, the expected Lagrangian density for $\bm{L}$ becomes
\begin{align}
\mathcal{L}_{L}=&\frac{\hbar^2}{2ZJa^3} \left[(1-\langle \pi^{a}\pi^{a}\rangle ^{{\rm th}})\partial_{t}\bm{L}\cdot\partial_{t}\bm{L} \right. \notag\\
&\quad \left. +\mathscr{A}_{ca}\mathscr{A}_{cb}\langle \pi^{a}\pi^{b}\rangle ^{{\rm th}}\right] +\frac{K}{2a^{3}}\left(\bm{L}\cdot\bm{e}_{\parallel}\right)^{2}. \label{eq:eff}
\end{align}
The key issue boils down to the calculation of the thermal correlation function $\left\langle \pi^{a}\pi^{b}\right\rangle ^{{\rm th}}$. To fulfill this task, we perform a Wick rotation for the $\bm{\pi}$ field and freeze the time variable of $\bm{L}$. Since deviations of $\bm{L}$ from $\bm{e}_{\parallel}$ are small and $\bm{\pi}$ is virtually perpendicular to $\bm{e}_{\parallel}$, we have $\left(\bm{\pi}\cdot\bm{e}_{\parallel}\right)^{2}\ll|\bm{\pi}|^{2}\left(\bm{L}\cdot\bm{e}_{\parallel}\right)^{2}$. Consequently, we can ignore the $\left(\bm{\pi}\cdot\bm{e}_{\parallel}\right)^{2}$ term in Eq.~\eqref{eq:Lpi}. In the non-interacting order, $\bm{\pi}$ is just a Klein-Gordon field with dispersion $\omega_{k}=\sqrt{c^2k^2+ZJK/\hbar^2}$. The Matsubara propagator is
\begin{align}
&\left\langle \pi^{a}\left(\tau,\bm{r}\right)\pi^{b}\left(0,0\right)\right\rangle _{0}^{{\rm th}} \notag\\
&\qquad = \frac{ZJa^3}{\beta\hbar^2}\delta^{ab} \sum_{n} \int\frac{{\rm d}^{3}\bm{k}}{\left(2\pi\right)^{3}}e^{i\left(\bm{k}\cdot\bm{r}+\omega_{k}\tau\right)}\frac{1}{\omega_{n}^{2}/\hbar^{2}+\omega_{k}^{2}}\nonumber \notag\\
&\qquad = \frac{ZJa^3}{2}\delta^{ab}\int\frac{{\rm d}^{3}\bm{k}}{\left(2\pi\right)^{3}}\frac{e^{i\bm{k}\cdot\bm{r}}}{\hbar\omega_{k}}\left[f_{B}\left(\omega_{k}\right)e^{\omega_{k}\tau}\right. \notag\\
&\qquad\qquad\qquad\qquad \left.+\left(1+f_{B}\left(\omega_{k}\right)\right)e^{-\omega_{k}\tau}\right],
\end{align}
where $\tau$ is the imaginary time, $f_{B}\left(\omega_{k}\right)=\frac{1}{e^{\beta\omega_{k}}-1}$ with $\beta=1/k_{B}T$, and $\omega_{n}=\frac{2\pi n}{\beta}$ ($n\in\mathbb{Z}$) is the Matsubara frequency. A relevant quantity that can be constructed from the propagator is the one-loop integral
\begin{align}
&\left\langle \pi^{a}\left(\tau,\bm{r}\right)\pi^{b}\left(\tau,\bm{r}\right)\right\rangle ^{{\rm th}} \notag\\
&\qquad =\frac{Za^3}{2}\delta^{ab}\int\frac{{\rm d}^{3}\bm{k}}{\left(2\pi\right)^{3}}\frac{J}{\hbar\omega_{k}}\coth\left(\frac{\beta\hbar\omega_{k}}{2}\right). \label{eq:loop}
\end{align}
Following the same spirit, terms in Eq.~\eqref{eq:odd} that are odd in the power of the $\bm{\pi}$ field should vanish identically: $\left\langle |\bm{\pi}|^{2n}\pi^{a}\right\rangle ^{{\rm th}}=\left\langle |\bm{\pi}|^{2n}\pi^{a}\bm{\pi}\cdot\partial_{t}\bm{\pi}\right\rangle ^{{\rm th}}=\left\langle |\bm{\pi}|^{2n}\partial_{t}\pi^{a}\right\rangle ^{{\rm th}}=0$. Moreover, $\left\langle \pi^{b}\partial_{t}\pi^{a}\right\rangle ^{{\rm th}}\sim\int{\rm d}\omega_{k}\omega_{k}\int{\rm d}^{3}\bm{k}\left\langle \pi^{b}\pi^{a}\right\rangle ^{{\rm th}}$, which vanishes as well. Therefore, there is no term in $\mathcal{L}_{{\rm odd}}$ that can survive the thermal averaging process.

Finally, inserting Eq.~\eqref{eq:loop} into Eq.~\eqref{eq:eff} and noticing that $\mathscr{A}_{ca}\mathscr{A}_{cb}\delta^{ab}=\dot{\bm{L}}^2$, we arrive at
\begin{align}
\mathcal{L}_{L}=\frac{\hbar^2}{2ZJa^3}\Theta(T) \dot{\bm{L}}^{2}+\frac{K}{2a^3}\left(\bm{L}\cdot\bm{e}_{\parallel}\right)^{2},\label{eq:conclusion}\\
\Theta(T)=1-\frac{Za^3}{\hbar} \int\frac{\mathrm{d}^3\bm{k}}{(2\pi)^3} \frac{J}{\omega_k}\coth\frac{\hbar\omega_k}{2k_BT}. \label{eq:theta}
\end{align}
If $\bm{k}$ is quantized due to geometric confinement, then $\int\frac{{\rm d}^{3}\bm{k}}{\left(2\pi\right)^{3}}$ should be understood as $\frac{1}{V}\sum_{\bm{k}}$. Equations~\eqref{eq:conclusion} and~\eqref{eq:theta} prove our central result, Eqs.~\eqref{eq:Leff} and~\eqref{eq:IT}.

\section{}\label{nA}

As expressed by Eq.~\eqref{eq:alphabeta}, $\alpha(t)$ and $\beta(t)$ parametrize the deviation of $\bm{L}(t)$ from $\bm{e}_{\parallel}(t)$ in the body frame, while $\gamma$ is a constant. Since the coordinates in the body frame are time dependent, we first need to relate them to the laboratory frame coordinates
\begin{widetext}
\begin{align}
\left(\begin{array}{c}
\bm{e}_{1}(t)\\
\bm{e}_{2}(t)\\
\bm{e}_{3}(t)
\end{array}\right)=\left(\begin{array}{ccc}
\cos\phi\cos\psi-\cos\theta\sin\phi\sin\psi & \sin\phi\cos\psi+\cos\theta\cos\phi\sin\psi & \sin\theta\sin\psi\\
-\cos\phi\sin\psi-\cos\theta\sin\phi\cos\psi & -\sin\phi\sin\psi+\cos\theta\cos\phi\cos\psi & \sin\theta\cos\psi\\
\sin\theta\sin\phi & -\sin\theta\cos\phi & \cos\theta
\end{array}\right)\left(\begin{array}{c}
\bm{e}_{x}\\
\bm{e}_{y}\\
\bm{e}_{z}
\end{array}\right),
\end{align}
\end{widetext}
where all Euler angles depend on time. Next we insert the above expression into Eq.~\eqref{eq:alphabeta}, followed by inserting thus-obtained $\bm{L}(t)$ into Eq.~\eqref{eq:conclusion}. Expanding Eq.~\eqref{eq:conclusion} to quadratic orders in $\alpha$ and $\beta$, we have
\begin{widetext}
 \begin{align}
 \mathcal{L}_{L}&= \frac{\hbar^2}{2ZJa^3}\Theta(T)g_{ij}\dot{\lambda}^{i}\dot{\lambda}^{j} +\frac{\hbar^2}{2ZJa^3}\Theta(T)\left(\dot{\alpha}^{2}+\sin^{2}\gamma\dot{\beta}^{2}\right) +\frac{K}{a^{3}}\left(1-\alpha^{2}-\beta^{2}\sin^{2}\gamma\right)\nonumber \\
 &\qquad\qquad + \frac{\hbar^2}{ZJa^3}\Theta(T)\left\{ \dot{\alpha}\left(-\sin\psi\dot{\theta}+\cos\psi\sin\theta\dot{\phi}\right)\right. \notag\\
 &\qquad\qquad\qquad\qquad \left. +\dot{\beta}\sin\gamma\left[\sin\gamma\left(\cos\theta\dot{\phi}+\dot{\psi}\right) -\cos\gamma\left(\cos\psi\dot{\theta}+\sin\theta\sin\psi\dot{\phi}\right)\right]\right\} ,\label{eq:Int}
 \end{align}
\end{widetext}
where the leading term $g_{ij}$ is the spin-mechanical inertia that does not depend on $\alpha$ and $\beta$.

As the adiabatic approximation has frozen the time for the Euler angles, the integral over $\bm{L}$ converts into that over $\alpha$ and $\beta$, which is Gaussian type according to Eq.~\eqref{eq:Int}. The measure of the integral is
\begin{align}
{\rm d}^{3}\bm{L}\delta\left(\bm{L}^{2}-1\right) & =\left|\begin{array}{cc}
\frac{\partial L_{3}}{\partial\alpha} & \frac{\partial L_{3}}{\partial\beta}\\
\frac{\partial}{\partial\alpha}\arctan\frac{L_{2}}{L_{1}} & \frac{\partial}{\partial\beta}\arctan\frac{L_{2}}{L_{1}}
\end{array}\right|{\rm d}\alpha{\rm d}\beta \notag\\
&=\sin\gamma{\rm d}\alpha{\rm d}\beta.
\end{align}
To perform the integral, we set
\begin{align}
 x_1=&\alpha,\\
 x_2=&\beta\sin\gamma, \\
 X_{1}=&-\sin\psi\dot{\theta}+\cos\psi\sin\theta\dot{\phi}, \\
 X_{2}=&\sin\gamma(\cos\theta\dot{\phi}+\dot{\psi}) \notag\\
 &\quad-\cos\gamma(\cos\psi\dot{\theta}+\sin\theta\sin\psi\dot{\phi}).
\end{align}
Then we finally obtain
\begin{widetext}
 \begin{align}
 \mathcal{Z}_{AF}= & \int\mathcal{D}x_{1}\mathcal{D}x_{2}\exp\left\{ i\frac{\hbar V\Theta(T)}{ZJa^3}\int{\rm d}t \left[\frac{1}{2}\left(\dot{x}_{1}^{2}+\dot{x}_{2}^{2}\right)-\frac{ZJK}{\hbar^2\Theta(T)}\left(x_{1}^{2}+x_{2}^{2}\right)+\left(\dot{x}_{1}X_{1}+\dot{x}_{2}X_{2}\right)\right]\right\} \nonumber \\
 = & \int\mathcal{D}x_{1}\mathcal{D}x_{2}\exp\left\{ i\frac{\hbar V\Theta(T)}{2ZJa^3} \int{\rm d}t\left[x_{1}\left(\partial_{t}^{2}- \frac{2ZJK}{\hbar^2\Theta(T)}\right)x_{1}+x_{2}\left(\partial_{t}^{2}-\frac{2ZJK}{\hbar^2\Theta(T)}\right)x_{2}-2\left(x_{1}\dot{X}_{1}+x_{2}\dot{X}_{2}\right)\right]\right\} \nonumber \\
 = & \frac{(V/a^3)\sqrt{2K/(ZJ)}\Theta(T)}{2\pi i\sin\left[ \sqrt{\frac{2ZJK}{\hbar^2\Theta(T)}}\left(t_{f}-t_{i}\right) \right]} \exp\left\{\frac{i}{\hbar}\sum_{n=1}^{\infty}\frac{\hbar^{2}V\Theta(T)}{2ZJa^{3}}\int\mathrm{d}tX^{{\rm T}}\left[-\frac{\hbar^{2}\Theta(T)}{ZJK}\partial_{t}^{2}\right]^{n}X \right\} ,
 \end{align}
\end{widetext}
where $X=\left\{ X_{1},X_{2}\right\} $, and $t_{f}$ ($t_{i}$) is the upper (lower) limit of the time integral.


\begin{thebibliography}{20}
\bibitem{note_anisotropy} The full form of the anisotropy should be $\left| \bm{M}\cdot\bm{e}_{\parallel}\right|^2+\left| \bm{N}\cdot\bm{e}_{\parallel}\right|^2$ where $\bm{M}$ is the magnetization vector. But in deriving the effective action of $\bm{N}$, \textit{e.g.}, see Ref.~\onlinecite{ref:Fradkin}, the first term is negligible since $K\ll J$.
\bibitem{note_temp} The temperature-dependent factor $\Theta(T)$ is derived in the adiabatic limit where the staggered field is in thermal equilibrium, so using it here for the non-adiabatic correction is not exact, but rather an approximation.
\end{thebibliography}
\end{document}